\documentclass[preprint2,twoside]{hwo}

\newcommand{\hst}{{\it HST}}
\newcommand{\bfhst}{\emph{HST}}

\newcommand{\jwst}{{\it JWST}}
\newcommand{\hwo}{{\it HWO}}
\newcommand{\tess}{{\it TESS}}
\newcommand{\xmm}{{\it XMM-Newton}}

\newcommand{\msol}{M$_{\odot}$}
\newcommand{\rearth}{R$_{\oplus}$}

\newcommand{\lya}{Ly$\alpha$}

\input{hwo.h}

\begin{document}

\title{{\bf \LARGE The unique ability of the \emph{Hubble Space Telescope} to characterize young exoplanet environments}}

\author{
Keighley~E.~Rockcliffe,$^{1,2}$
Allison~Youngblood,$^{2}$
Kevin~France,$^{3}$
Cynthia~Froning,$^{4}$
P.\,C.~Schneider,$^{5}$
Elisabeth~Newton,$^{6}$
David~J.~Wilson,$^{3}$
Vighnesh~Nagpal,$^{7,8}$
Sarah~Peacock,$^{1,2}$
Seth~Redfield,$^{9}$ 
Mayumi~Liz de~Andrade Miyazato,$^{10}$
Hans-R.~M\"uller,$^{6}$
Aylin~Garcia~Soto,$^{11}$
}

\affil{$^1$\small\it University of Maryland Baltimore County; $^2$\small\it NASA Goddard Space Flight Center; $^3$\small\it University of Colorado; $^4$\small\it Southwest Research Institute; $^5$\small\it Kiel University; $^6$\small\it Dartmouth College; $^7$\small\it University of Chicago; $^8$\small\it NSF Graduate Research Fellow; $^9$\small\it Wesleyan University; $^{10}$\small\it Colorado State University; $^{11}$\small\it Boise State University \vspace{-1.5ex}}

\author{\footnotesize{\bf Corresponding author:} keighley.e.rockcliffe@nasa.gov \vspace{-3ex}}

\author{\footnotesize{\bf Endorsed by:} 
Adina D. Feinstein (Michigan State University), Shreyas Vissapragada (Carnegie Science Observatories), Leonardo A. Dos Santos (STScI), Ava Morrissey (Carnegie Science Observatories), R.\,O. Parke Loyd (Eureka Scientific)}

\begin{abstract}
The chemical and mass evolution of exoplanet atmospheres is shaped by their specific X-ray through ultraviolet ($5 - 3200$ \AA) irradiance history. X-ray and EUV ($5 - 911$ \AA) radiation largely contributes to atmospheric heating via photoionization, while far- and near-UV emission ($912 - 3200$ \AA) drives photochemistry. The (uncharacterized) variance between young star spectra in this wavelength range for the same spectral type causes significant uncertainty in interpreting present-day transmission spectra of young exoplanets, directly impacting the science return of the {\it James Webb Space Telescope} and other programs. Additionally, the lack of direct X-ray through UV characterization for stars of all ages leads to large uncertainties in the high-energy irradiance history of all planetary systems, propagating into uncertainties in their chemical and mass evolution. This influences current and future observing programs, as well as the goal of the future flagship {\it Habitable Worlds Observatory} to find and characterize habitable exoplanets. There are less than a handful of young planet hosts with well-characterized X-ray through UV environments. The {\it Hubble Space Telescope} is the only observatory capable of measuring the UV spectrum and enabling the characterization of exoplanet high-energy environments. We advocate for an observing program to measure the UV, estimate the EUV, and measure the X-ray where possible and needed of all amenable young planet hosts, addressing the Space Telescope Science Institute's call for {\it Building a Roadmap for Hubble Science into the 2030s}.
\\
\\
\end{abstract}

\section{X-ray, EUV, and UV-driven atmospheric processes}
Stars of spectral type FGKM (the most common exoplanet hosts) have strong magnetic fields, which cause powerful high-energy emission from the X-ray, extreme ultraviolet, and ultraviolet (Table~\ref{tab:spec}). That high-energy emission photoionizes and photodissociates atmospheric species for the exoplanets they host, injecting heat and regulating chemical changes and mass loss \citep{MiguelExploring2014,WangDusty2019,GaoDeflating2020}. Stellar X-ray to UV radiation changes dramatically and is the strongest at young ages. For example, young Sun-like stars can output 100 - 1000 times stronger X-ray and EUV, 10 - 60 times stronger FUV, and 10 - 20 times stronger NUV emission than the modern Sun \citep{RibasEvolution2005,ShkolnikHAZMAT2014}. M dwarfs show a few orders of magnitude decay in X-ray, FUV, and NUV luminosity between 10 Myr and about 3 Gyr old, with the X-ray exhibiting a steeper decline \citep{2013MNRAS.431.2063S}. This evolution in stellar high-energy emission over a few hundred million years after formation in particular drives drastic atmospheric chemical, thermal, and mass evolution on exoplanets.

\begin{table}[!]
    \centering
    \caption{Relevant wavelength regions for high-energy-driven exoplanet evolution.} \label{tab:spec}
    \begin{tabular}{|c|c|}
    \hline
    Region & Wavelength Range (\AA) \\
    \hline
    Ultraviolet (UV) & $912 - 3200$ \\
    Near-ultraviolet (NUV) & $1700 - 3200$ \\
    Far-ultraviolet (FUV) & $912 - 1700$ \\
    Lyman-$\alpha$ (\lya) & $1215.67$ \\
    \hline
    X-ray and EUV (XUV) & $1 - 911$ \\
    Extreme ultraviolet (EUV) & $101 - 911$ \\
    X-ray & $5 - 100$ \\
    \hline
    \end{tabular}
\end{table}

\begin{figure*}[ht]
    \centering
    \includegraphics[width=0.95\linewidth]{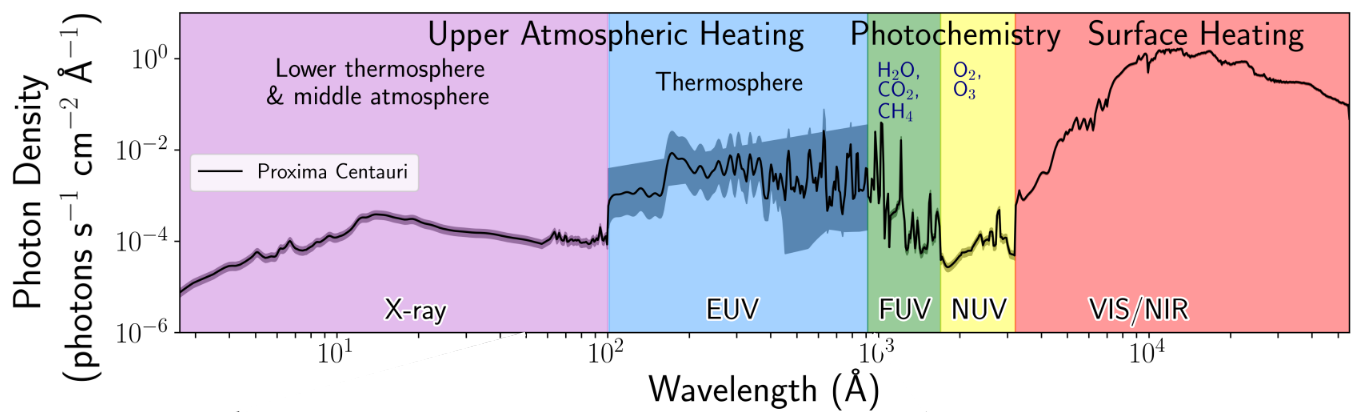}
    \caption{Modified Figure 1 from \cite{2022JATIS...8a4006F}. The composite spectrum of Proxima Centauri depicting the different wavelength regions (roughly corresponding to Table~\ref{tab:spec}) and how they influence exoplanet atmospheres.}
    \label{fig:high-energy}
\end{figure*}

\vspace{-6pt}
\subsection{Atmospheric escape} \label{sec:escape}

Stellar radiation at wavelengths shortward of $< 1100$ \AA\ photoionizes atoms and molecules in the upper layers of an exoplanet's atmosphere, depositing heat that drives escape \citep{Tian2008}. The UV spectrum from \hst\ is vital for reconstructing the currently unobservable EUV that dominates heating the upper layers of atmospheres (\S\ref{sec:obs}). Young volatile-rich exoplanets with short orbital periods experience more irradiation and are even more vulnerable to escape. The mass evolution of these exoplanets due to escape has significantly shaped current exoplanet compositions and demographics \citep[e.g.,][]{Szabo2011,Lopez2012,Fulton2017,Owen2018b}. A planet's specific high-energy irradiance history is a key factor in atmospheric retention, and it follows that the large uncertainty in their nascent X-ray through UV environment, especially the EUV, leads to large uncertainty in their mass evolution.

\begin{figure*}[ht]
    \centering
    \includegraphics[width=0.85\linewidth]{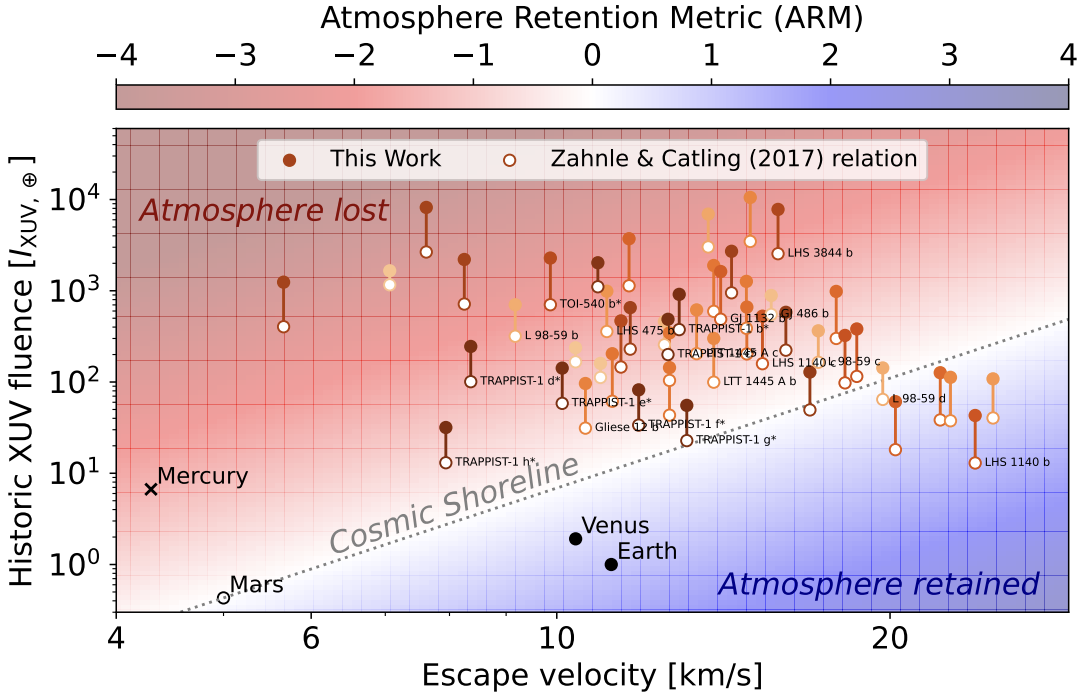}
    \caption{Modified \cite{2025ApJ...986L...3P} Figure 2 showing an updated Cosmic Shoreline for exoplanets orbiting full convective M dwarfs. The exoplanets shift significantly along the y-axis depending on their assumed XUV history (specifically showing \citealt{2025ApJ...986L...3P} and \citealt{2017ApJ...843..122Z}), which changes the location of the Cosmic Shoreline.}
    \label{fig:shore}
\end{figure*}

The ``Cosmic Shoreline" is an empirically-defined boundary in the integrated XUV irradiance - gravitational potential plane of Solar System planets that has been applied to separate exoplanets with retained atmospheres from those without \citep{2017ApJ...843..122Z}. This boundary is a widely used metric for constructing the most scientifically impactful high-priority targets for atmospheric characterization. Because of this and the large uncertainty in its y-axis, XUV irradiance, the Cosmic Shoreline has been revised many times \citep{2025ApJ...986L...3P,2025ApJ...992..198J,2025arXiv250702136B,2025arXiv250812865M}. \cite{2025ApJ...986L...3P}, for example, reconstructed the Cosmic Shoreline considering the elongated period of high XUV irradiance experienced by exoplanets that orbit fully convective M dwarfs (Figure~\ref{fig:shore}). 
Their {\it assumed} integrated XUV irradiance differs from exoplanets that orbit higher mass M dwarfs and changes the location of the Cosmic Shoreline. This and similar work are compromised by the lack of knowledge of XUV histories for all stars. XUV measurements at a range of ages (and masses) where stellar high-energy output changes the most ($< 1$ Gyr) would allow for greater certainty in the Cosmic Shoreline, directly enabling the efficient use of current observatories, such as the {\it James Webb Space Telescope} (\jwst), and the preparation of high-priority targets for the {\it Habitable Worlds Observatory} (\hwo).

Photoionizing radiation also regulates whether an escaping neutral hydrogen atmosphere is observable. Lyman-$\alpha$ (\lya; $1215.67$ \AA) transits are a cornerstone of escape studies, directly probing neutral hydrogen as it escapes the planet and accelerates to high velocities \citep{Vidal-Madjar2003,Ehrenreich2015}. Analyses of \lya\ transits have been used to estimate mass loss rates, the speed and geometry of the escaping material, and constraints on the interacting stellar wind \citep[e.g.,][]{Ehrenreich2015,Bourrier2016,Esquivel2019,Khodachenko2019,Shaikhislamov2021}. Open questions regarding atmospheric escape and its influence on the exoplanet population motivate the \hst\ Multi-Cycle Treasury Survey STEL$\alpha$ (GO-17804; PIs: Loyd, Vissapragada), the largest exoplanet program awarded by \hst\ to date, and its goal to obtain dozens of \lya\ transits. The {\it time-dependent} high-energy irradiation of each planet governs whether its neutral hydrogen outflow will remain neutral and observable at \lya\ \citep{Rockcliffe2021,2023MNRAS.518.4357O}. This is especially true for young exoplanets experiencing elevated and significantly variable high-energy irradiation. While the majority of STEL$\alpha$ targets are older, better X-ray through UV constraints for young planet hosts will be necessary to plan future escape programs that probe larger parameter spaces, including the Earth-like planet regime discussed in \hwo\ Science Case: {\it Exoplanet Atmospheric Escape Observations with the Habitable Worlds Observatory} \citep{2025arXiv250707124D,2025JATIS..11d2236D}.

\vspace{-6pt}
\subsection{Photochemistry}

\begin{figure}[!]
    \centering
    \includegraphics[width=0.49\textwidth]{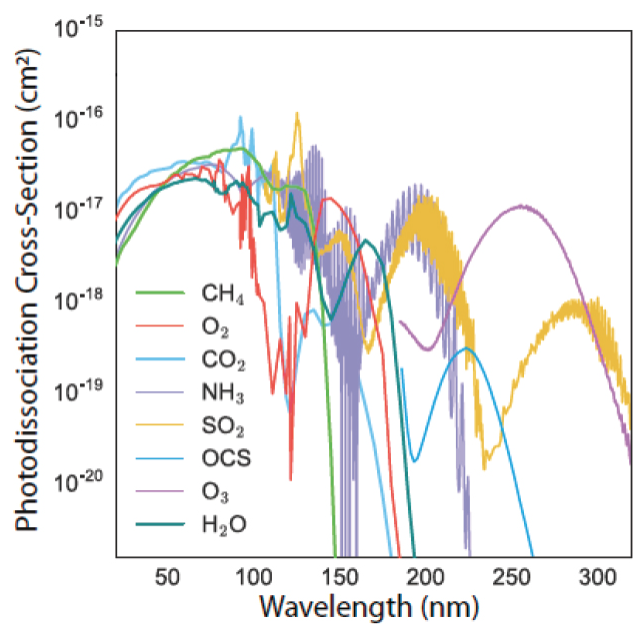}
    \caption{Figure 4.12 from \cite{NAP25187}. The photoabsorption cross sections of common atmospheric molecules are large and highly wavelength dependent throughout the UV.}
    \label{fig:mols}
\end{figure}

The UV continuum and emission lines photodissociate common molecules (Figure~\ref{fig:mols}), altering the the chemical makeup of an atmosphere \citep{HuPHOTOCHEMISTRY2012,MiguelExploring2014,TianHigh2014,LoydMUSCLES2016}. For the majority of exoplanet hosts (M$_{\star}$ $<2$ \msol), Lyman-$\alpha$ radiation dominates the UV and significantly contributes to chemistry in exoplanet atmospheres, dissociating molecules of interest like H$_2$O and CH$_4$ and governing the interpretation of potential biosignatures \citep{2015ApJ...809...57R}. \cite{TealEffects2022} investigated the sensitivity of photochemical models of Earth-like planet atmospheres orbiting M dwarfs to stellar UV spectrum input. They concluded  that ``fully observing a host star's UV spectrum, including multiple UV emission lines and the underlying continuum, remains the gold standard for modeling exoplanet atmospheres.''

The discovery of planets outside of the Solar System begs the question of whether extrasolar life exists as well. This question drives exoplanet characterization with \jwst, as well as the goal to discover and characterize habitable worlds with \hwo. Many intricate exoplanetary system properties must align to enable life as we know it. The X-ray through UV radiation from an exoplanet's host star must be weak enough to allow atmospheric retention and leave critical molecular components unharmed by dissociation. However, the UV emission must be strong enough to drive key processes related to life and the prebiotic chemistry responsible for its origin \citep[e.g.,][]{Mulkidjanian03,Sarker13,Rapf16}. This balance between destruction and creation will be finely tuned to each system depending on its high-energy environment, already motivating the X-ray through UV characterization of \hwo\ Target Stars and Systems \citep{2025AJ....170..293P}.

\vspace{-6pt}
\section{The UV is needed to interpret young planet transmission spectra}

\hst\ and now \jwst's capabilities for transmission spectroscopy in the infrared, which probes atmospheric composition, facilitate molecular abundance measurements and/or the identification of cloud decks and hazes. The transmission spectroscopy of young exoplanets yet to be altered by evolution offer a unique view on the processes that shape exoplanet demographics at large:
\begin{enumerate} \vspace{-5pt}
\item The exoplanets' atmospheres are near-primordial, having experienced minimal mass loss and chemical evolution. This means that the atmospheric composition will more closely track where the planet formed in the disk, e.g. high C/O ratios indicating formation beyond the snow line \citep{ObergEffects2011}. \vspace{-6pt}
\item The exoplanets provide the means to test different hypotheses for the trends seen in the atmospheres of older exoplanets. For example, the strength of the water absorption feature seen in \hst\ transmission spectra suggests a parabolic dependence on temperature, but the mechanisms behind this remain uncertain \citep{BrandeClouds2024}. \vspace{-6pt}
\item Observations of young exoplanets can provide a window into the compositions of super-Earths and sub-Neptunes. At older ages, these exoplanets are often characterized by either high mean molecular weight atmospheres or clouds \citep{BrandeClouds2024}, which inhibit studying their atmospheric abundances with transmission spectroscopy. Due to their different entropies, ages, and incident radiation, these same exoplanets at younger ages could have clear(er) atmospheres \citep[e.g.,][]{2024NatAs...8..899B}. \vspace{-6pt}
\item Transmission spectroscopy provides the means to measure the mass of exoplanets (in the case of a gas-rich planet only; \citealt{BatalhaChallenges2017}) due to the dependence of atmospheric scale height on planetary surface gravity \citep{deWitConstraining2013}. This is particularly useful for young exoplanets, where surface gravities may be very low due to high entropy \citep{LopezUNDERSTANDING2014} and mass measurements from radial velocities are challenging due to stellar activity \citep[e.g.,][]{BluntOverfitting2023}. \vspace{-6pt}
\end{enumerate}

Because of their unique and powerful window into gaps in our understanding of exoplanets, young exoplanets have received substantial investment of observational resources. For example, there are around 34 programs awarded across \jwst's Cycles 1 - 5 devoted to young planet characterization, with 130 hours awarded to GO-05959 (PI: Feinstein) alone to constrain young sub-Neptune compositions. 

Knowledge of the X-ray through UV radiation received by these exoplanets is needed in order to accurately interpret their infrared transmission spectroscopy, exemplified by HIP 67522 b: HIP 67522 b is an inflated planet ($10$ \rearth) orbiting a Sun-like star in the Sco-Cen OB association, meaning it has a reliable age measurement of 17 Myr \citep{RizzutoTESS2020,2024ApJ...973L..30B}. The planet's upper atmosphere is likely dominated by H$_2$ and is, therefore, vulnerable to photodissociation. The dissociation puffs up the atmosphere by halving the mean molecular weight for atmospheric pressures $<1$ $\mu$bar. This decreases the amplitudes of spectral features that are formed at these low pressures (specifically, H$_2$O and CO$_2$) as those molecules are subject to photodissociation themselves. Thus, the planet's UV irradiation is mimicking a larger planetary mass. A weak SO$_2$ feature was also detected, produced through the photodissociation of H$_2$S. HIP 67522's high-energy spectrum was critical in the interpretation of its \jwst\ transmission spectrum and constraining the planet's true mass (Figure 10; \citealt{2024AJ....168..297T}).

\vspace{-6pt}
\section{The status of young host X-ray through UV characterization}

For the aforementioned reasons and more, the community desires X-ray through UV spectral characterization of planet hosts \citep{2019BAAS...51c.408C,2019BAAS...51c.522L}. The age-, activity-, and mass-dependent stellar high-energy spectrum originates from the chromosphere, transition region, and corona. Unlike the photosphere, the emission from stellar atmospheres is not predictable from fundamental stellar properties.

Decaying relationships between activity (often X-ray luminosity) and stellar rotation or time are used as inputs for exoplanet evolution studies to understand how the current exoplanet population came to be \citep[e.g.,][]{2021A&A...656A.157B,2022A&A...661A..23F,2026ApJ..1001..133G}. There are significant assumptions baked into these inputs, neglecting diverging spin-down histories and the uncertainty in converting X-ray to EUV and UV evolution, which propagate into uncertainties in the evolution of exoplanets. For example, \cite{2021A&A...649A..96J} explored stellar rotation, X-ray, EUV, and \lya\ radiation across stellar properties and highlighted the need for a more comprehensive treatment of stellar evolution than single activity decay relations. Young stars, in particular, could experience fast or slow spin-down evolution, widening the uncertainty in their activity levels and subsequent high-energy output at a given age \citep{TuExtreme2015,2021A&A...649A..96J,KetzerInfluence2023}.

The MUSCLES Treasury Survey and its successors pioneered measuring the high-energy environment of exoplanets, providing the community with dozens of well-characterized panchromatic spectra for planet hosts \citep{2016ApJ...820...89F,2019ApJ...871L..26F,2025ApJ...978...85W}. However, these surveys focus on old field stars. The FUMES and HAZMAT surveys both focus on young M dwarfs in the far-ultraviolet (FUV; $1000 - 1700$ \AA), but do not contain planet hosts or higher mass stars \citep{2021ApJ...911..111P,2014AJ....148...64S}. The young low-mass stars targeted by ULLYSES were explicitly T Tauri stars \citep{Roman-DuvalOverview2020}, which are all highly active, i.e., these data do not cover the subsequent decay in magnetic activity relevant for the evolution of planetary atmospheres.

There are some scaling relations between the UV and longer wavelength activity indicators \citep[e.g.,][]{MelbourneEstimating2020}. These activity-based relations may differ for young stars depending on the specific part of the stellar atmosphere they probe and the physical mechanism behind the relation's saturation \citep{LinskyRelative2020}. Regardless, direct UV characterization greatly reduces uncertainty in exoplanet abundance measurements from atmospheric retrievals (e.g., hazy Earth-like atmospheres; \citealt{TealEffects2022}).

\vspace{-6pt}
\section{A major \bfhst\ initiative to fill this crucial gap}

The lack of young planet host high-energy spectra is a bottleneck for exoplanet science. \hst\ can fill this gap relatively efficiently while leveraging its existing synergy with {\it XMM-Newton}. Currently, there are of order 30 young planet hosts with incomplete X-ray through UV characterization, including a handful of key \jwst\ targets. 

\vspace{-6pt}
\subsection{The specific capabilities of \hst\ required} \label{sec:obs}

\hst's Space Telescope Imaging Spectrograph (STIS) covers the UV at sufficient sensitivity and resolution for this initiative. STIS's low resolution first-order G140L ($1150 - 1736$ \AA) and G230L ($1570 - 3180$ \AA) gratings cover the FUV emission lines and NUV continuum, respectively. A few bright stars may require echelle gratings for their NUV continuum, with which \hst\ is equipped.

The stellar \lya\ emission line is heavily attenuated by neutral hydrogen and deuterium in the interstellar medium (ISM), as well as contaminated by Earth's airglow. The intrinsic stellar line must be reconstructed using the available observed signal, a model of the local ISM absorption, and a model of the stellar emission \citep{WoodStellar2005,YoungbloodMUSCLES2016}. The detection and reconstruction of each star's \lya\ emission will require \hst's medium resolution G140M ($1194 - 1249$ \AA) grating or echelle gratings (E140M or E140H) with STIS\footnote{Except for the brightest targets, \hst's Cosmic Origins Spectrograph (COS) saturates with Earth's airglow and \lya\ is unobtainable \citep{2023ApJ...946...98C}.}. The low resolution gratings do not always provide sufficient delineation of spectral features for confident reconstruction \citep{2026AJ....171...25F}. The unique capabilities of \hst\ are required, namely high resolving power at \lya, in order to make robust measurements of this important FUV line.

We advocate for contemporaneous X-ray and UV observations where possible and where there is no existing archival data, which requires the scheduling collaboration currently capable by \hst\ and \xmm. Non-contemporaneous observations in these two wavelength regimes may probe different activity states of each target, introducing uncertainty into the combined X-ray and UV spectrum. Additionally, the contemporaneous X-ray and FUV are used to estimate the currently unobservable EUV (due to lack of observatory coverage). This wavelength region can be reconstructed using the differential emission measure (DEM) technique \citep{DuvvuriReconstructing2021,2025ApJ...993..138D}. The DEM technique models the temperature and plasma density of the stellar atmosphere from the observed X-ray and UV emission line fluxes to infer excitation and emission for EUV emission lines.

\vspace{-6pt}
\subsection{Summary of impact}

The present moment is a confluence of several motivations for an \hst\ program to completely characterize the high-energy radiation from young planet hosts:
\begin{enumerate} \vspace{-5pt}
    \item The number of amenable targets is plateauing since the {\it Transiting Exoplanet Survey Satellite} (\tess) has completed its all-sky survey of nearby, bright stars, slowing the detection of exoplanets at astronomically precise young ages. \vspace{-6pt}
    \item Several targets recently have or will soon have \jwst\ transmission spectra, which need contemporaneous high-energy data. The more time passes between the \jwst\ and the X-ray to UV observations, the larger the uncertainty in each planet's high-energy environment at the time of those \jwst\ observations. This is due to uncharacterized stellar activity cycles which can introduce factor of two flux uncertainties. \vspace{-6pt}
    \item We still have access to \hst's unique UV capabilities! Roughly half of stellar high-energy emission remains uncertain without FUV data \citep{LinskyRelative2020}, which contributes significantly to photochemistry within exoplanet atmospheres and informs the highly uncertain EUV. The NUV continuum and emission lines are also crucial to interpreting and modeling photochemistry. \vspace{-6pt}
    \item We are in an important stage of \hwo\ science case development and observatory preparation. Knowledge of the high-energy irradiance history of present-day exoplanets will increase certainty in the atmospheric retention and characterization of potential targets, directly supporting \hwo's primary mission to search for and characterize habitable exoplanets. \vspace{-6pt}
\end{enumerate}

\newpage

\bibliography{references.bib}

\end{document}